\documentclass[aps,superscriptaddress,showpacs,a4paper]{revtex4} 
\usepackage{bm,dsfont}
\usepackage[dvips]{graphicx}

\newcommand{\qb}{\mathbf{q}}
\newcommand{\kb}{\mathbf{k}}
\newcommand{\xb}{\mathbf{x}}
\newcommand{\ub}{\mathbf{u}}
\newcommand{\tb}{\mathbf{t}}
\newcommand{\Rb}{\mathbf{R}}
\newcommand{\rb}{\mathbf{r}}

\begin{document}

\title{A Gaussian Model for the Membrane of Red Blood
Cells with Cytoskeletal Defects}

\author{Cyril Dubus}
\affiliation{Laboratoire Mati\`ere et Syst\`emes Complexes (MSC), 
UMR 7057 CNRS \& Universit\'e Paris~7, 2 place Jussieu, F-75251 Paris
Cedex 05, France}
\affiliation{Laboratoire de Physico-Chimie Th\'eorique, UMR CNRS 7083 \&
ESPCI, 10 rue Vauquelin, F-75231 Paris Cedex 05, France}
\author{Jean-Baptiste Fournier}
\thanks{author for correspondance}
\affiliation{Laboratoire Mati\`ere et Syst\`emes Complexes (MSC), 
UMR 7057 CNRS \& Universit\'e Paris~7, 2 place Jussieu, F-75251 Paris
Cedex 05, France}
\affiliation{Laboratoire de Physico-Chimie Th\'eorique, UMR CNRS 7083 \&
ESPCI, 10 rue Vauquelin, F-75231 Paris Cedex 05, France}

\begin{abstract}
We study a Gaussian model of the membrane of red blood cells: a
``phantom" triangular network of springs attached at its vertices to a
fluid bilayer with curvature elasticity and tension. We calculate its
fluctuation spectrum and we discuss the different regimes and
non-monotonic features, including the precise crossover at the mesh size
between the already known limits with two different tensions and the
renormalisation of the bending rigidity at low wavevectors. We also show
that the non-diagonal correlations reveal, in ``dark field",
the cytoskeletal defects. As a first step toward a non-invasive defect
spectroscopy, the specific case of lacking bonds is studied numerically
and analytically.  
\end{abstract}

\date{\today}
\maketitle


The outer walls of biological cells are fluid bilayer membranes made by
the self-assembly of lipid molecules in water~\cite{alberts_book}. For
about three decades, the physical properties of these two-dimensional
fluid surfaces, including elasticity, thermal fluctuations, phase
behavior, topology, shape polymorphism, etc., have been thoroughly
investigated in model systems~\cite{nelson_book,safran_book,%
mouritsen_book}.  In real cells, membranes host a large number of
proteins and other inclusions, and they are attached to a network of
filaments called the cytoskeleton~\cite{alberts_book}. In most cases
this network is three-dimensional. In the case of red blood cells (RBC's),
however, the cytoskeleton is a triangular two-dimensional network of
flexible spectrin filaments attached to the membrane by proteins located
at its vertices~\cite{byers85,bennett89,mohandas94}. 
Recent experimental and theoretical studies have focused on the shape
elasticity~\cite{boal98,boal_review,lim02,mukho02,lenormand01} and
fluctuation properties~\cite{zilker87,gov03,fournier04,gov04,gov05} of
these composite membranes. Up to now, the available theoretical models
describing the RBC membrane fluctuations either are
mean-field~\cite{fournier04} or describe the cytoskeleton as a fixed
plane producing a checkerboard (square) modulated
potential~\cite{gov04}. In order to obtain a better description, it is
essential to release the assumption of a fixed cytoskeletal plane and to
allow for coupled fluctuations of the membrane and the cytoskeleton. It
is also desirable to take into account the triangular nature of the
cytoskeleton and to determine the effects of the cytoskeletal
defects~\cite{seung88,gov05}. 

\begin{figure}
\centerline{\includegraphics[width=.6\textwidth]{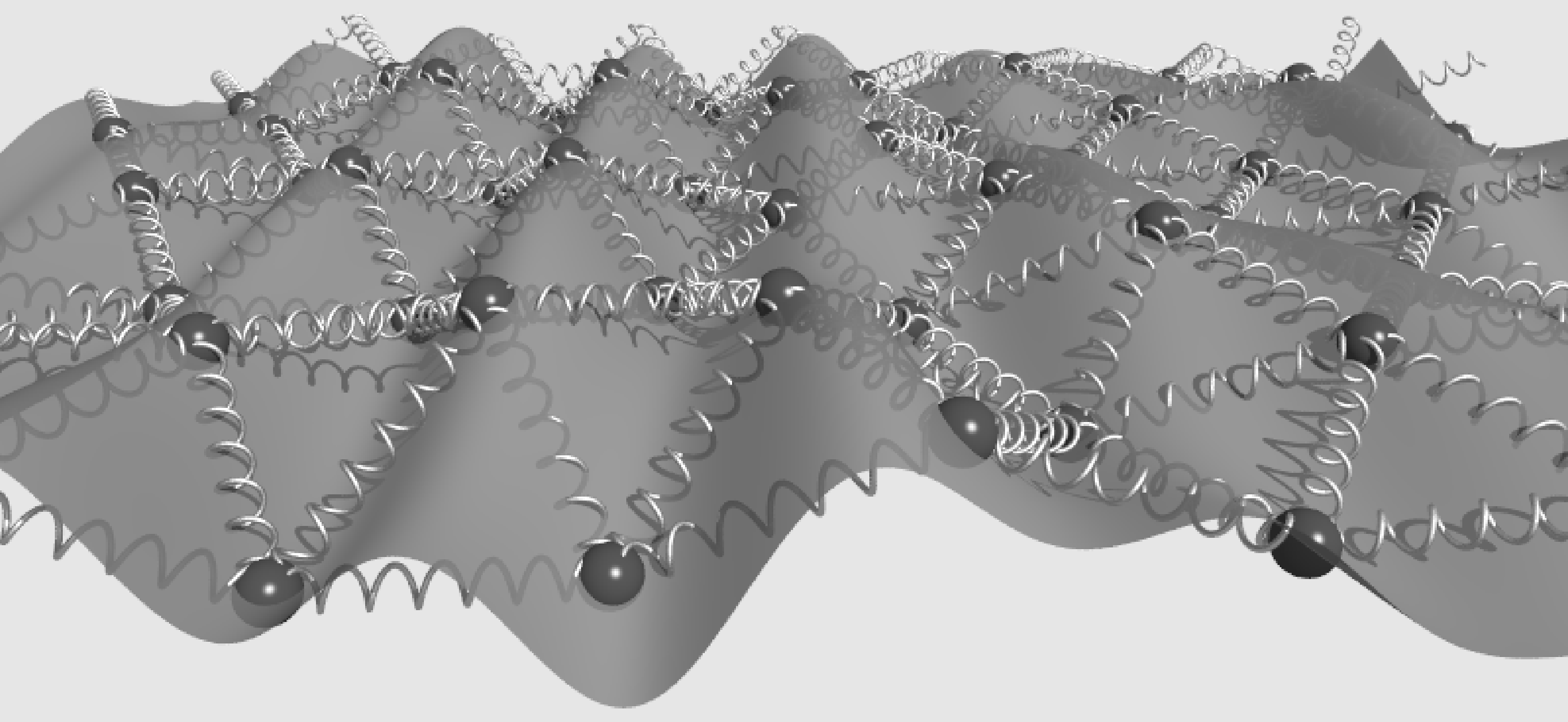}}
\caption{Sketch of the model red blood cell (RBC) membrane.}
\label{art}
\end{figure}

In this paper, we propose a simple gaussian model describing the
composite membrane of RBC's. It allows to calculate explicitely the
whole fluctuation spectrum, even in the presence of cytoskeletal
defects, such as lacking bonds and seven-fold or five-fold
defects. We model the composite membrane (Fig.~\ref{art}) as a
triangular network of harmonic springs attached at its vertices
to a fluid membrane described by a curvature Hamiltonian plus
surface tension~\cite{helfrich73}.  The only coupling between the two
subsystems is the constraint that the cytoskeleton vertices lie within
the membrane surface. As in model polymers~\cite{degennes_book}, the
gaussian character arises from neglecting the excluded volume between
the membrane and the cytoskeleton (thereby treating the latter as
``phantom"~\cite{Kantor87}).  Depending on the experimental conditions,
RBC fluctuations may be of thermal origin, or they may be driven by ATP
consumption~\cite{tuvia97,tuvia98}.  Here, we consider only the thermal,
equilibrium situation.  We derive the height fluctuation
spectrum $\langle |h_\qb|^2\rangle$ of our model system, and we discuss
the crossover between the different regimes and the non-monotonic
features arising from the membrane--cytoskeleton coupling. Since the
cytoskeleton and its defects break the translational symmetry, we study
the extinction conditions for the non-diagonal spectrum, i.e., $\langle
h_\qb h_\kb\rangle$ for $\kb\ne-\qb$.  We demonstrate that the latter
reveals, in ``dark field", the cytoskeletal defects. Detail studies
should therefore allow to determine the nature and the density of the
defects. Here, we discuss preliminary results concerning the spectral
signature of lacking spectrin bonds. 

\begin{figure}
\centerline{\includegraphics[width=.5\textwidth]{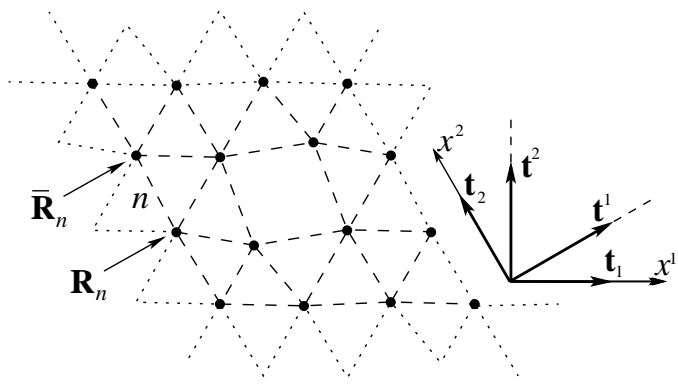}}
\caption{Network of $N$ springs (dashed lines) in a $M\times M$ unit
cell of a cytoskeleton with periodic boundary conditions (dots) along
the basis vectors $\tb_1$ and $\tb_2$. Here, $M=3$ and $N=26$. The
extremities of the $n$th spring have projections $\Rb_n$ and
$\overline{\Rb}_n$ onto the reference plane. A bond is lacking, in
order to illustrate a simple situation with a density of defects.}
\label{cyto}
\end{figure}

Because of the triangular symmetry of the cytoskeleton, we introduce a
planar reference frame with unitary basis vectors $\tb_1$ and $\tb_2$
making an angle of $2\pi/3$ (Fig.~\ref{cyto}). We define the
dual basis $(\tb^1,\tb^2)$ by
$\tb_\alpha\cdot\tb^\beta=\delta_\alpha^\beta$, where
$\delta_\alpha^\beta$ is the Kronecker symbol. The metric tensor
$g_{\alpha\beta}=\tb_\alpha\cdot\tb_\beta$ has diagonal (resp.\
off-diagonal) elements equal to $1$ (resp.\ $-\frac{1}{2}$) and
determinant $g=\mathrm{det}(g_{\alpha\beta})=\frac{3}{4}$. Vectors in
the plane will be represented by their coordinates according to
$\xb=x_\alpha\,\tb^\alpha=x^\alpha\,\tb_\alpha$ (Einstein summation
assumed throughout), where $x^\alpha=g^{\alpha\beta}x_\beta$ with
$g^{\alpha\beta}=g_{\alpha\beta}^{-1}$. We describe the membrane by its
elevation $h(\xb)\equiv h(x^\alpha)$ above a reference plane parallel to
$(\tb_1,\tb_2)$ and we assume periodic boundary conditions such that
$h(x^1,x^2)=h(x^1+L_1,x^2)=h(x^1,x^2+L_2)$, $\forall x^\alpha$. Our model
Hamiltonian (per unit cell) is the following:
\begin{equation}
\label{H}
\mathcal{H}=\int_0^{L_1}\!\!\!\int_0^{L_2}\!\!\!\!
\sqrt{g}\,dx^1dx^2 
\left[\frac{\kappa}{2}\left(\partial_\alpha\partial^\alpha
h\right)^2+\frac{\sigma}{2}(\partial^\alpha h)(\partial_\alpha h)
\right]
+\sum_{n=1}^N\frac{\mu_n}{2}
\left[h(R_n^\alpha)-h(\bar R_n^\alpha)\right]^2.
\end{equation}
The first term, $\mathcal{H}_0[h]$, is the standard membrane Helfrich
Hamiltonian~\cite{helfrich73} expressed in our non-orthonormal basis,
with $\kappa$ the membrane bending rigidity and $\sigma$ an externally
imposed surface tension. The second term, $\mathcal{V}[h]$, modeling the
elasticity of the spectrin network, represents the stretching energy
associated with the $N$ springs that lie within the repeat period. In
this expression, $R_n^\alpha$ and $\bar R_n^\alpha$ are the projections
onto the reference plane of the extremities of the $n$th spring
(Fig~\ref{cyto}).  We take for $R_n^\alpha$ and $\bar R_n^\alpha$ the
equilibrium positions that they would occupy in the case the membrane
were flat, since in the limit of small Gaussian fluctuations the
in-plane and the out-of-plane fluctuations of the vertices are
decoupled. We allow for any kind of defect or heterogeneity, therefore
we do not assume that $R_n^\alpha$ and $\bar R_n^\alpha$ make a regular
triangular network.  The constant $\mu_n$ is related to the rigidity $k$
and free-length $\ell$ of the springs (which we assume to be identical
biological elements) and to the geometry in the following way: to second
order in $h$, the spring elongation energy
$\frac{1}{2}k[\{(\Rb_n-\overline{\Rb}_n)^2+
[h(\Rb_n)-h(\overline{\Rb}_n)]^2\}^{1/2}-\ell]^2$ gives the in-plane
contribution $\frac{1}{2}k(|\Rb_n-\overline{\Rb}_n|-\ell)^2$ plus the
term $\mathcal{V}[h]$ of Eq.~(\ref{H}), with
\begin{equation}
\mu_n=k\left(1-\ell|\Rb_n-\overline{\Rb}_n|^{-1}\right).
\end{equation}
Because the network may have been stretched or compressed, either
mechanically or osmotically, or because of the possible defects, the
$\mu_n$ may have either sign.  

Let us define a Fourier transform by
$h(x^1,x^2)=(\sqrt{g}L_1L_2)^{-1/2}\sum_\qb h_\qb\,e^{iq_\alpha
x^\alpha}$ where the wavevectors are quantified as
$q_\alpha=n_\alpha2\pi/L_\alpha$, where $n_\alpha\in\mathds{Z}$.
In order to work with dimensionless quantities, we normalize the
energies by $k_\mathrm{B}T\equiv\beta^{-1}$, the in-plane lengths
by $\xi_\sigma=\sqrt{\kappa/\sigma}$ and the membrane height
$h(x^\alpha)$ by $1/\sqrt{\beta\sigma}$ . Hence $h_\qb$ is normalized by
$\sqrt{\kappa/\beta}\,\sigma^{-1}$ and $\mu_n$ by $\sigma$. Unless otherwise
specified, all quantities are now dimensionless. We thus have
$\mathcal{H}_0=\sum_{\qb,\kb}\frac{1}{2}h_\qb G^{-1}_{\qb,\kb}h_\kb$
where $G_{\qb,\kb}=\delta_{\qb+\kb}/(\qb^2+\qb^4)$ and
$\qb^2=q^\alpha q_\alpha$, yielding the correlation function of the bare
membrane:
\begin{equation}
G(\xb)=\langle
h(\mathbf{0})h(\xb)\rangle=
\frac{1}{\sqrt{g}L_1L_2}\sum_\qb\frac{e^{iq_\alpha x^\alpha}}
{\qb^2+\qb^4}\,.
\end{equation}
For an infinite membrane, this gives a Bessel function of the second
kind:
$G(\xb)=-\frac{1}{4}\mathrm{Y_0}(|\xb|)$.
Because $\mathcal{H}$ is not diagonal in $\qb$-space due to the
dependence in $\Rb_n$ and $\overline{\Rb}_n$, it
cannot be directly inverted to yield the full correlation
function $G(\xb)=\langle h(\mathbf{0})h(\xb)\rangle$. To proceed, we add
an external field, $\mathcal{H}\to\mathcal{H}+\sum_\qb h_\qb
f_{-\qb}$, and we calculate $\langle h_\qb h_\kb
\rangle=-\partial^2F/(\partial f_{-\qb}\partial f_{-\kb})|_{f=0}$, where
$F=-\ln[\int\mathcal{D}[h]\exp(-\mathcal{H})]$ is the free-energy. In
order to integrate the membrane degrees of freedom, we rewrite
$\exp(-\mathcal{V})$ with the help of auxialiary fields (one for each
spring) as
\begin{equation}
\int\!\prod_nd\phi_n\,\exp
\sum_n\left(-\frac{\phi_n^2}{2\mu_n}
+i\phi_n[h(R_n^\alpha)-h(\bar R_n^\alpha)]
\right).
\end{equation}
Sums and products over $n$ (and further on $m$) implicitely run from 1
to $N$.  Performing the gaussian integral over $h$ yields
$\exp(-F)=\int\!\prod_nd\phi_n\,\exp(-\tilde{\mathcal{H}})$, where 
\begin{equation}
\tilde{\mathcal{H}}=\sum_n\frac{\phi_n^2}{2\mu_n}-\frac{1}{2}\sum_{\qb,\kb}
(S_{-\qb}\!-\!f_{-\qb})\,G_{\qb,\kb}\,(S_{-\kb}\!-\!f_{-\kb})\,,
\end{equation}
in which
$S_\qb=(\sqrt{g}L_1L_2)^{-\frac{1}{2}}\sum_ni\phi_n(
e^{-iq_\alpha R_n^\alpha}-e^{-iq_\alpha \bar R_n^\alpha})$.
The integral over the $\phi_n$ is again Gaussian, yielding
$F[f]=-\frac{1}{2}\sum_{\qb,\kb}f_{-\qb}G_{\qb,\kb}f_{-\kb}+\delta F[f]$,
with
\begin{eqnarray}
\delta F[f]\!&=&\!\frac{1}{2}\sum_{n,m}
\left(\sum_\qb c_\qb^{(n)}f_{-\qb}\right)
B_{nm}^{-1}
\left(\sum_\kb c_\kb^{(n)}f_{-\kb}\right),~
\end{eqnarray}
where
\begin{eqnarray}
c_\qb^{(n)}\!&=&\!\frac{1}{(\sqrt{g}L_1L_2)^\frac{1}{2}}
\frac{e^{-iq_\alpha R_n^\alpha}-e^{-iq_\alpha \bar R_n^\alpha}}
{\qb^2+\qb^4} \,,\\
B_{nm}&=&\frac{1}{\mu_n}\delta_{nm}+
G(\Rb_n\!-\!\Rb_m)-G(\Rb_n\!-\!\overline{\Rb}_m)
-G(\overline{\Rb}_n\!-\!\Rb_m)
+G(\overline{\Rb}_n\!-\!\overline{\Rb}_m)\,.
\end{eqnarray}
It follows that the full correlation function $\Gamma_{\qb,\kb}=\langle
h_\qb h_\kb\rangle=G_{\qb,\kb}+\Delta\Gamma_{\qb,\kb}$ is given, with no approximation, by
\begin{equation}
\label{res}
\Gamma_{\qb,\kb}=\frac{\delta_{\qb+\kb}}{\qb^2+\qb^4}
-\sum_{n,m=1}^N
c_\qb^{(n)}\,B_{nm}^{-1}\,c_\kb^{(m)}.
\end{equation}
For weak cytoskeletal strains, i.e.,
$\mu_n=\mathcal{O}(\mu)\ll1$, we have 
$B_{nm}^{-1}=\mu_n\delta_{nm}+\mathcal{O}(\mu^2)$, and thus
\begin{equation} 
\label{res2}
\Delta\Gamma_{\qb,\kb}\simeq-\frac{1}{\sqrt{g}L_1L_2}
\sum_{n=1}^N\mu_n
\frac{\psi_{n,\qb}\,\psi_{n,\kb}}{(\qb^2+\qb^4)(\kb^2+\kb^4)}
\end{equation} where
$\psi_{n,\qb}=e^{-i\qb\cdot\Rb_n}-e^{-i\qb\cdot\overline{\Rb}_n}$.
The fluctuation spectrum can thus be obtained analytically for weak
strains even in presence of defects, or, for
larger strains, numerically by simply inverting a
$N\times N$ matrix.

\medskip\textit{Uniform cytoskeleton}.~--~~In the case of a uniform
dilation or compression in the absence of defects, the points $\Rb_n$ and
$\overline{\Rb}_n$ form a regular triangular lattice with $\mu_n=\mu$,
$\forall n$, corresponding to a uniform meshsize $\xi$ such that
$N=3L_1L_2/\xi^2$. Thus, in 
dimensional units and for small values of $\mu/\sigma$, Eq.~(\ref{res2})
gives for $\kb=-\qb$ the following fluctuation spectrum:
\begin{equation}
\label{spectrum_homo}
\langle|h_\qb|^2\rangle\simeq
\frac{k_\mathrm{B}T}{\sigma\,\qb^2+\kappa\,\qb^4}
-\frac{4\mu\, g^{-\frac{1}{2}}\xi^{-2}\, k_\mathrm{B}T}
{\left(\sigma\,\qb^2+\kappa\,\qb^4\right)^2}
\sum_{j=0}^2\sin^2\!\left[\frac{q\xi}{2}
\cos(\theta\!-\!\frac{j\pi}{3})\right]\!,~~~
\end{equation}
with $\theta$ the angle between $\tb_1$ and $\qb$.
For $q\ll\xi^{-1}$, we get 
$k_\mathrm{B}T/\langle|h_\qb|^2\rangle\simeq
\sigma_\mathrm{eff}\,q^2+\kappa_\mathrm{eff}\,q^4$, with
\begin{eqnarray}
\sigma_\mathrm{eff}&\simeq&\sigma+\frac{3}{2\sqrt{g}}\,\mu
~=~\sigma+\frac{k}{\sqrt{3}}\,(1-\ell/\xi),\\
\kappa_\mathrm{eff}&\simeq&\kappa-\frac{3}{32\sqrt{g}}\,\mu\xi^2
~=~\kappa+\frac{\sqrt{3}}{16}\,k\xi\left(\xi-\ell\right)\,,
\end{eqnarray}
and for $q\gg\xi^{-1}$, Eq.~(\ref{spectrum_homo}) yields the
bare membrane spectrum $k_\mathrm{B}T/\langle|h_\qb|^2\rangle\simeq
\sigma\,q^2+\kappa\,q^4$. We thus recover the effective tension jump
$\sigma-\sigma_\mathrm{eff}$ discussed in Refs.~\cite{gov03,fournier04}.
In addition, we obtain the renormalization of the bending rigidity
$\kappa_\mathrm{eff}$ due to the cytoskeleton. Taking
$k\simeq10^{-5}$~J/m$^2$~\cite{lenormand01} and
$\xi\approx\ell\simeq10^{-7}$\,m~\cite{shen86}, we obtain
$\kappa_\mathrm{eff}-\kappa\simeq10^{-20}$\,J, wich is comparable but
smaller than the usual membrane bending rigidity, in agreement with the
measurements of Ref.~\cite{strey95}. 

\begin{figure} 
\centerline{\includegraphics[width=.5\textwidth]{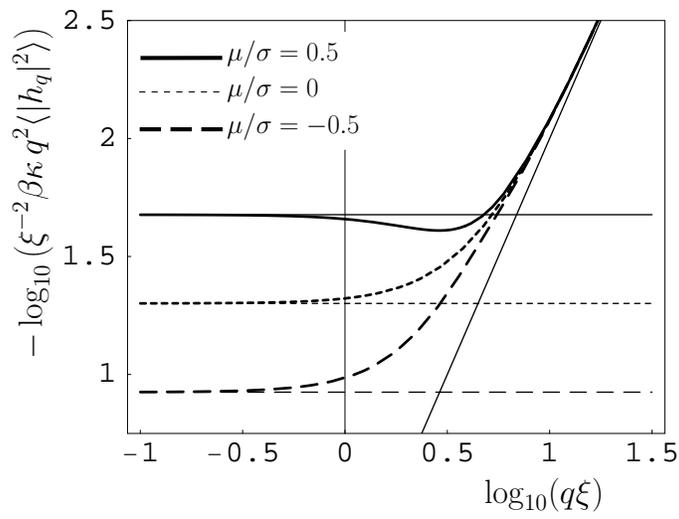}}
\caption{Fluctuation spectrum of the model RBC membrane with
a uniform triangular cytoskeleton, for a tension
corresponding to $\sqrt{\sigma/\kappa}\simeq4.5\,\xi^{-1}$ and a direction of
$\qb$ corresponding to $\theta=0$. The case $\mu>0$
(resp.\ $\mu<0$) corresponds to a stretched (resp.\ compressed)
cytoskeleton. We chose $|\mu|/\sigma=0.5$ to better emphasize the
features of the curves, although Eq.~(\protect\ref{spectrum_homo}) is
valid in the limit $\mu\ll\sigma$. All quantities are in dimensional
units.} 
\label{spectre} 
\end{figure}

The full fluctuation spectrum for large tensions, i.e.,
$\sigma>\kappa/\xi^2$, is shown in Fig.~\ref{spectre}.  There are three
regimes: i) a flat, tension-dominated regime for $q<\xi^{-1}$, revealing
$\sigma_\mathrm{eff}$, ii) a crossover region
$\xi^{-1}<q<\sqrt{\sigma/\kappa}$, showing the transition from
$\sigma_\mathrm{eff}$ to $\sigma$ and thus exhibiting a negative slope
for large enough $\mu>0$, iii) a terminal linear regime dominated
by the bending rigidity, revealing $\kappa$. In the case of weak
tensions, i.e., $\sigma<\kappa/\xi^2$ (not shown), the transition to the
terminal regime occurs before $q$ reaches $\xi^{-1}$, hence the change
in tension becomes invisible. Since $\kappa_\mathrm{eff}\simeq\kappa$,
one does not see either the slope change occuring around
$q\simeq\xi^{-1}$. The curves thus look like the dashed ones in
Fig.~\ref{spectre}, except that the intersection of the asymptotes
occurs in the region $q<\xi^{-1}$. 

\medskip \textit{Cytoskeleton with defects}.~--~~Let us discuss the
general conditions under which the couplings $\langle h_\qb
h_\kb\rangle$ are extinct or not. Our Hamiltonian can be expressed as
$\mathcal{H}=\int\!d^2x\,d^2r\,h(\xb)\,H(\xb,\rb)\,h(\xb+\rb)
=\int\!d^2q\,d^2k\,h_\qb\,\bar{H}_{\qb,\kb}\,h_\kb$. In the absence of a
cytoskeleton, the translational invariance of the system yields the
well-known non-extinction condition $\qb=-\kb$. In the presence of a
cytoskeleton, this invariance breaks down. For a perfectly uniform
cytoskeleton, however, $H(\xb,\rb)$ remains invariant for discrete
translations of the form $x^\alpha\to x^\alpha+\xi$  and the
non-extinction condition becomes $q_\alpha=-k_\alpha$ \textit{modulo
$2\pi/\xi$}. This weaker condition is still strong, because $2\pi/\xi$
is very large (macroscopically).  Now, in the presence of random
structural defects, the translational symmetry vanishes and all the
couplings $\langle h_\qb h_\kb\rangle$ appear~\cite{notearray}. 

\begin{figure}
\centerline{\includegraphics[width=.5\textwidth]{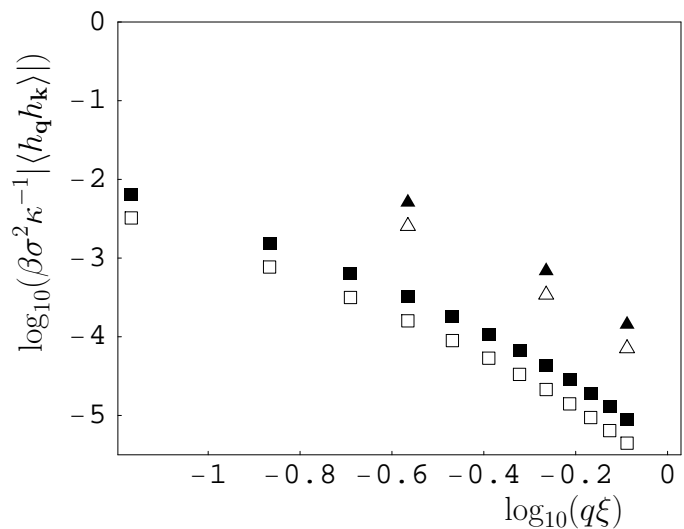}}
\caption{Non-diagonal fluctuation spectrum $\langle h_\qb h_\kb\rangle$ in the
presence of lacking bonds, for $\qb\ne-\kb$ but $\|\qb\|=\|\kb\|=q$, as
detailed in the text. The squares (resp.\ triangles) correspond to one
missing bond per unit cell, as in Fig.~\protect\ref{cyto}, but for
$M=80$ (resp.\ $M=20$). The values of $q$ are quantified by the
periodic boundary conditions. The filled symbols are numerically
calculated from Eq.~(\protect\ref{res2}), while the open symbols
correspond to the analytical approximation discussed in the text.}
\label{nondiag}
\end{figure}

There are mainly two types of defects: lacking bonds and dislocations
made by pairs of five-fold and seven-fold defects~\cite{seung88,gov05}.
A systematic study of the signature of these defects according to their
density and type is outside the scope of this paper~\cite{prepa}. Here,
we limit ourselves to the simplest situation, i.e., lacking bonds, as in
Fig.~\ref{cyto}. We show in Fig.~\ref{nondiag} the corresponding
non-diagonal spectrum (i.e., $\langle h_\qb h_\kb\rangle$ with
$\qb\ne-\kb$) for $\qb=q\,\tb^1/\|\tb^1\|$ and
$\kb=-q\,\tb^2/\|\tb^2\|$, as obtained from Eq.~(\ref{res2}) after
numerically minimizing the in-plane elastic energy in order to determine
the positions $R_n^\alpha$ and $\bar R_n^\alpha$ of the springs ends. A
satisfying analytical approximation of this spectrum
(Fig.~\ref{nondiag}) can be obtained by neglecting the lattice
distorsion, i.e., by using the positions $R_n^\alpha$ and $\bar
R_n^\alpha$ of the regular lattice while setting one of the $\mu_n$ to
zero. In this case, Eq.~(\ref{res2}) gives $|\langle h_\qb
h_\kb\rangle|=\mu\, k_\mathrm{B}Tg^{-\frac{1}{2}}L^{-2}
|\sin(\qb\cdot\rb)\sin(\kb\cdot\rb)|(\sigma q^2+\kappa q^4)^{-2}$, with
$\rb=\frac{1}{2}\xi\ub$ where $\ub$ is the unit vector parallel to the
missing bond. In the $q\to0$ limit, we obtain
\begin{equation}
|\langle h_\qb h_\kb\rangle|\approx
\frac{\mu\,k_\mathrm{B}T}{\sigma^2 q^2}\,\rho\,,
\end{equation}
which shows that the non-diagonal spectrum is proportional to the defect
density $\rho=1/(3M^2)$.

In conclusion, in the light of this model experimental measurements of
RBC equilibrium fluctuations under strain could give new informations
concerning the structure and elasticity of the cytoskeleton. Since the
defects and the plasticity of RBC's are believed to be controled by
the ATPase activity~\cite{gov05}, it should also be interesting
to investigate the (equilibrium) non-diagonal spectrum at various stages
of the erythrocyte life or in particular conditions, such as capillary
transport. Finally, it should be possible to study in a Langevin picture
the dynamical properties of this model system, thereby determining the
consequence of non-equilibrium ATP-driven noise on the fluctuation
spectrum. This would be extremely interesting, as recent experiments
reveal strong non-equilibrium effects~\cite{tuvia97,tuvia98,gov05}.

\acknowledgments

We thank Dr.\ David Lacoste for stimulating discussions at the early
stage of this work and Prof.\ Ken Sekimoto for precious comments and
fuitfull discussions all along.



\begin{thebibliography}{99}
\bibitem{notearray}For an array of defects, however, there will be
(weak) extinction conditions due to the translational symmetry.
\end{thebibliography}
\end{document}